\documentclass[twocolumn,twoside,slac_two]{revtex4}
\pdfoutput=1
\usepackage{graphicx}
\usepackage{fancyhdr}
\begin{document}

\title{The GAMMA-400 Space Mission}

\author{P. Cumani}
\affiliation{Istituto Nazionale di Fisica Nucleare, Sezione di Trieste and Physics Department of University of Trieste, Trieste, Italy}
\author{A.M. Galper, N.P. Topchiev, V.A. Dogiel,	Yu.V. Gusakov, S.I. Suchkov}
\affiliation{Lebedev Physical Institute, Russian Academy of Sciences, Moscow, Russia}

\author{I.V. Arkhangelskaja, A.I. Arkhangelskiy,	G.L. Dedenko, V.V. Kadilin, V.A. Kaplin, A.A. Kaplun, M.D. Kheymits, A.A. Leonov, V.A. Loginov, V.V. Mikhailov,	P.Yu. Naumov, M.F. Runtso, A.A. Taraskin, E.M. Tyurin,	 Yu.T. Yurkin, V.G. Zverev}
\affiliation{National Research Nuclear University MEPhI, Moscow, Russia}

\author{S.G. Bobkov, M.S. Gorbunov, A.V. Popov, O.V. Serdin}
\affiliation{Scientific Research Institute for System Analysis, Russian Academy of Sciences, Moscow, Russia}

\author{R.L. Aptekar, E.A. Bogomolov, M.V. Ulanov, G.I. Vasilyev}
\affiliation{Ioffe Institute, Russian Academy of Sciences, St. Petersburg, Russia}

\author{A.L. Men'shenin}
\affiliation{Research Institute for Electromechanics, Istra, Moscow region, Russia}

\author{V.N. Zarikashvili}
\affiliation{Pushkov Institute of Terrestrial Magnetism, Ionosphere, and Radiowave Propagation, Troitsk, Moscow region, Russia}

\author{V. Bonvicini, M. Boezio, F. Longo, E. Mocchiutti, A. Vacchi, N. Zampa}
\affiliation{Istituto Nazionale di Fisica Nucleare, Sezione di Trieste and Physics Department of University of Trieste, Trieste, Italy}

\author{O. Adriani, E. Berti, M. Bongi, S. Bottai, N. Mori, P. Papini, P. Spillantini, A. Tiberio, E. Vannuccini}
\affiliation{Istituto Nazionale di Fisica Nucleare, Sezione di Firenze and Physics Department of University of Florence, Firenze, Italy}

\author{G. Castellini, S. Ricciarini}
\affiliation{Istituto di Fisica Applicata Nello Carrara - CNR and Istituto Nazionale di Fisica Nucleare, Sezione di Firenze, Firenze, Italy}

\author{G. Bigongiari, S. Bonechi, P. Maestro, P.S. Marrocchesi}
\affiliation{Department of Physical Sciences, Earth and Environment, University of Siena and Istituto Nazionale di Fisica Nucleare, Sezione di Pisa, Italy}

\author{P.W. Cattaneo, A. Rappoldi}
\affiliation{Istituto Nazionale di Fisica Nucleare, Sezione di Pavia, Pavia, Italy}

\author{C. De Donato, P. Picozza, R. Sparvoli}
\affiliation{Istituto Nazionale di Fisica Nucleare, Sezione di Roma 2 and Physics Department of University of Rome Tor Vergata, Rome, Italy}

\author{M. Tavani}
\affiliation{Istituto Nazionale di Astrofisica IASF and Physics Department of University of Rome Tor Vergata, Rome, Italy}

\author{L. Bergstr{\"o}m}
\affiliation{Stockholm University, Department of Physics; and the Oskar Klein Centre, AlbaNova University Center, Stockholm, Sweden}

\author{	J. Larsson, M. Pearce, F. Ryde}
\affiliation{KTH Royal Institute of Technology, Department of Physics; and the Oskar Klein Centre, AlbaNova University Center, Stockholm, Sweden}

\author{A.A. Moiseev}
\affiliation{CRESST/GSFC and University of Maryland, College Park, Maryland, USA}

\author{I.V. Moskalenko}
\affiliation{Hansen Experimental Physics Laboratory and Kavli Institute for Particle Astrophysics and Cosmology, Stanford University, Stanford, USA}

\author{B.I. Hnatyk}
\affiliation{Taras Shevchenko National University of Kyiv, Kyiv, Ukraine}
\author{V.E. Korepanov}
\affiliation{Lviv Center of Institute of Space Research, Lviv, Ukraine}

\begin{abstract}
GAMMA-400 is a new space mission which will be installed on board the Russian space platform Navigator. It is scheduled to be launched at the beginning of the next decade. GAMMA-400 is designed to study simultaneously gamma rays (up to 3 TeV) and cosmic rays (electrons and positrons from 1 GeV to 20 TeV, nuclei up to 10$^{15}$-10$^{16}$ eV). Being a dual-purpose mission, GAMMA-400 will be able to address some of the most impelling science topics, such as search for signatures of dark matter, cosmic-rays origin and propagation, and the nature of transients. GAMMA-400 will try to solve the unanswered questions on these topics by high-precision measurements of the Galactic and extragalactic gamma-ray sources, Galactic and extragalactic diffuse emission and the spectra of cosmic-ray electrons + positrons and nuclei, thanks to excellent energy and angular resolutions.
\end{abstract}

\maketitle

\thispagestyle{fancy}

\section{INTRODUCTION}
GAMMA-400 (\cite{2013arXiv1306.6175G}) is a Russian space mission, approved by the Russian space agency, with an international contribution. Foreseen to be launched at the beginning of the next decade, the satellite will be positioned on a circular orbit at $\sim$200000 km. This specific orbit, combined with a pointing mode observational strategy, allows to perform continuous observations of a source without Earth occultation. During its first year of mission, GAMMA-400 is planned to observe the Galactic plane.\\
Designed as a dual experiment, GAMMA-400 will be able to study gamma rays, from 100 MeV up to several TeV, as well as cosmic rays, electrons up to 20 TeV and protons and nuclei up to the ``knee'' (10$^{15}$-10$^{16}$ eV). It will search for possible dark matter signal thanks to an unprecedented energy resolution that will permit to detect features associated to dark matter annihilation or decay in the spectra of sources such as the Galactic Center. GAMMA-400 will also study gamma-ray sources such as active galactic nuclei, supernova remnants, pulsars and gamma-ray bursts (GRBs). The GRB study will be performed using both the main instrumentation, described in the next section, and the Konus-FG detectors. Six Konus-FG will be installed on GAMMA-400 to study GRBs in the 10 keV - 15 MeV energy range with a field-of-view of 2$\pi$ sr. Four of these detectors will be able to reconstruct the direction of the incoming photons with an accuracy between 0.5$^\circ$ and 3$^\circ$, while the remaining two will serve as spectrometric detectors.\\ 
GAMMA-400 will address the remaining issues regarding cosmic-rays origin, acceleration and propagation by studying the high energy all electron spectrum, with a 2\% energy resolution, and the cosmic-ray elemental spectra up to the knee, with high statistics and energy resolution.\\
Some of the GAMMA-400 performance, and the scientific objectives that will be addressed, are summarized in tab. \ref{t:scobj}.

\begin{table}[h]
\centering
\begin{small}
\begin{tabular}{ l c r}
Performance & & Scientific Objectives \\\hline
Energy Res. & $\sim 1$\% $\gamma$ & DM\\
 & $\sim 2$\% e$^\pm$ & CRs origin \\
 & $\sim 35$\% p & CRs propagation \\\hline
 Angular Res. & $\sim$0.6$^\circ$ @ 1 GeV & CRs origin \\
 & $\sim$0.02$^\circ$ @ 100 GeV & Transients \\
  & $\sim$0.006$^\circ$ @ 1 TeV & EBL \\\hline
  GF & $>3$ m$^2$sr & DM\\
 &  & CRs origin\\
  &  & CRs propagation\\\hline
\end{tabular}
\end{small} 
\caption{Summary of the GAMMA-400 performance and scientific topics.}
\label{t:scobj}
\end{table}

\section{GEOMETRY}

\begin{figure}[h]
\centering
\includegraphics[width=0.43\textwidth]{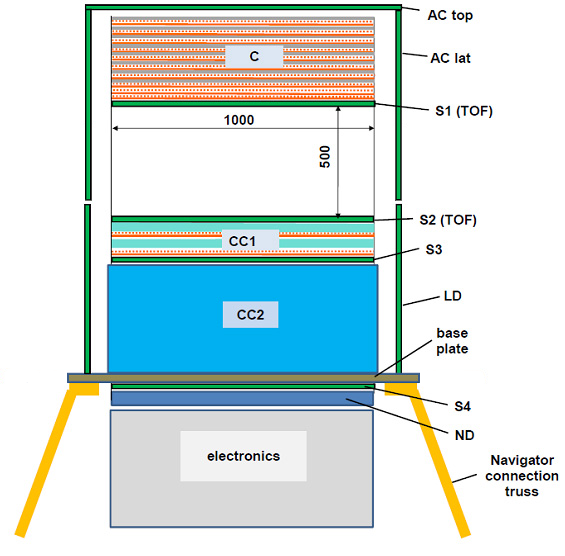}
\caption{Schematic view of the apparatus. From top to bottom: converter-tracker (C), anticoincidence system (AC top and lat), time-of-flight system (S 1 to 4), calorimeter (CC1 and CC2), charge identification system on the side of the calorimeter (LD) and neutron detector (ND). The dimensions values are in mm.} \label{baseline}
\end{figure}

The GAMMA-400 apparatus, of which a schematic view is presented in fig. \ref{baseline}, will comprise:
\begin{itemize}
\item A converter/tracker (C) where the impinging gamma ray creates an electron-positron pair subsequently detected by Silicon layers;
\item A calorimeter composed partially by CsI(Tl) slabs and Silicon sensors (CC1, also referred to as pre-shower in the following) and partly by CsI(Tl) cubes (CC2, also referred to as calorimeter in the following);
\item An Anticoincidence system covering both the sides and the top of the detector (AC top and lat) to reject the charged particles for gamma-ray observations;
\item A Time-of-flight system composed by four layers of scintillating materials (S1 and S2) to discriminate upgoing particles, such as backsplashed particles from the calorimeter, and downgoing particles;
\item A charge identification system (LD), to discriminate between the different elements interacting inside the detector;
\item A neutron detector (ND) and scintillation detectors (S3 and S4), used to improve the electron/hadron rejection factor.
\end{itemize}
A comparison between the tracker and calorimeter of \textit{Fermi} and GAMMA-400 is presented in tab. \ref{t:comp}.

\begin{table*}[t]
\centering
\begin{tabular}{l  c  c  c }
 & & Fermi & GAMMA-400\\\hline
Tracker & Dimension (cm$^2$) & $\sim$140$\times$140 & $\sim$97$\times$97\\\cline{2-4}
 & Radiation Length & 12 planes: 0.03 X$_0$ &  8 planes: 0.1 X$_0$ \\
 & & 4 planes: 0.18 X$_0$ & 2 planes: no W \\
  & & 2 planes: no W & \\\cline{2-4}
  & Pitch ($\mu$m) & 228 & 80 \\\cline{2-4}
  & Readout Pitch ($\mu$m) & 228 & 240 \\\cline{2-4}
 & Readout & Digital  & Analog\\\hline
Calorimeter & Vertical R.L. (X$_0$) & 8.6 & 23 (CC1 not included)\\\cline{2-4}
& Vertical I.L. ($\lambda_I$) & 0.4 & 1.1 (CC1 not included)\\\cline{2-4}
 & Segmentation & 96 Bars $\times$ Tower & 9408 Cubes \\
 & & 2.7$\times$2.0$\times$32.6 cm$^3$ & 3.6$\times$3.6$\times$3.6 cm$^3$ \\\hline    
\end{tabular}
\caption{Comparison between the tracker and calorimeter of \textit{Fermi} (\cite{2009ApJ...697.1071A}) and GAMMA-400}
\label{t:comp}
\end{table*}

\subsection{Tracker}

The tracker is divided into four towers. Each tower is composed by ten planes of single-sided Silicon detectors. The first eight planes are interleaved by a 0.1 X$_0$ of Tungsten, absent in the last two planes, for a total of $\sim$1 X$_0$ in the whole tracker. The tungsten, where present, and two Si layers, for the x and y view, are mounted on a honeycomb Al support. A 2 mm gap separates two different trays. 
Each Si layer is composed by an array of 5$\times$5 tiles each of which has a 9.7$\times$9.7 cm$^2$ area. Five tiles are wired-bonded together to form a $\sim$49 cm long ladder. The sensors are single-sided strip detectors with a strip pitch of 80 $\mu$m and a read-out pitch of 240 $\mu$m. The read-out of the strips is analog, similar to the one used by AGILE (\cite{2001AIPC..587..729T}). This read-out system permits to retain the information on the energy released inside the strip, allowing to reach a low error on the hit position (less than 40 $\mu$m) as well as using the tracker Si planes as a charge identifier. Thanks to this detector configuration, GAMMA-400 will be able to achieve an angular resolution at low energy comparable to the one of \textit{Fermi} front (\cite{2009ApJ...697.1071A}), as can be noticed at the left of fig. \ref{perf}, even with a more than doubled tungsten thickness.

\subsection{Pre-shower}
\begin{figure*}[t]
\centering
\includegraphics[width=0.49\textwidth]{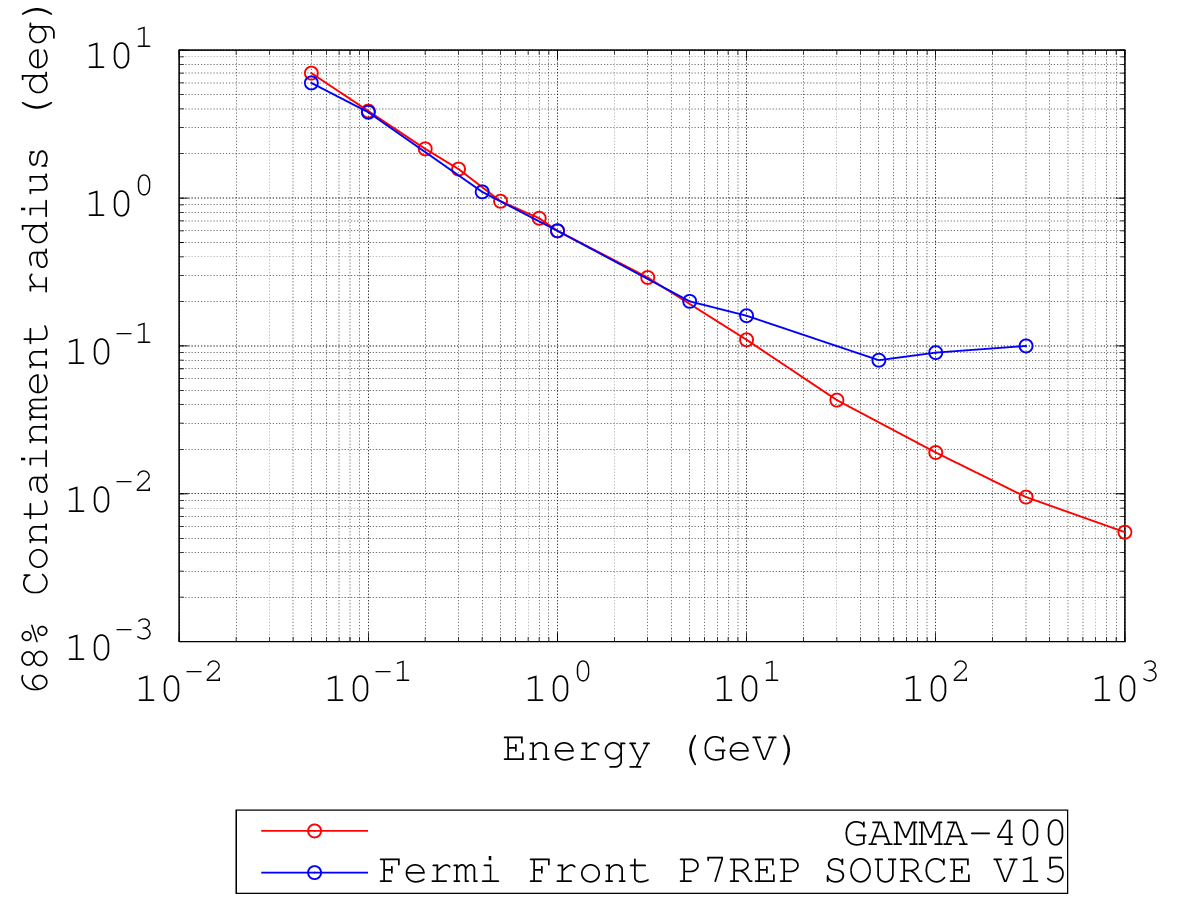}
\includegraphics[width=0.49\textwidth]{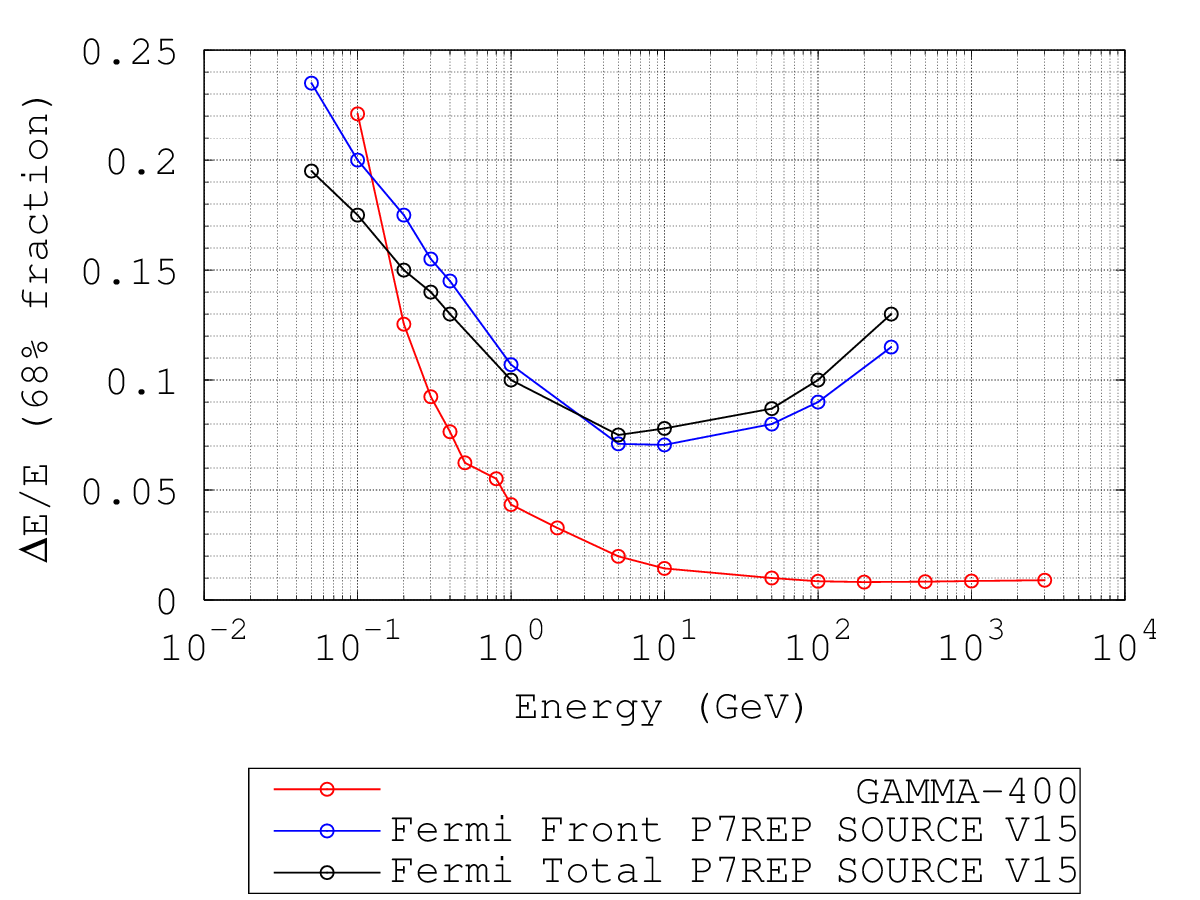}
\caption{Comparison between the angular (\textit{left}) and energy (\textit{right})
 resolutions of \textit{Fermi}-LAT (\cite{URL}) and GAMMA-400, using
  information from both calorimeter and tracker (\cite{2013arXiv1306.6175G}).}
\label{perf}
\end{figure*}

The pre-shower is composed by two planes of CsI(Tl) slabs interleaved by single-sided Si detectors, two layers for the x and y view. The Si layers are equal in pitch, dimensions and read-out to the Si layers inside the tracker. The total radiation length of the detector is of $\sim$ 2 X$_0$.\\
Each CsI(Tl) plane is divided in an array of 20$\times$3 slabs, each with a volume of 33.3$\times$5$\times$2 cm$^3$. The orientation of the slabs on the first plane is perpendicular to the orientation on the second plane in order to have the separate x and y view.\\
The 50 cm lever arm between the tracker and the pre-shower, combined with the finely pitched Si in both detectors, allows to reach an optimal angular resolution at high energy, as shown on the left of fig. \ref{perf}. A direction reconstruction can be also performed using only information from the pre-shower, helping in increasing the total effective area of the instrument.

\subsection{Calorimeter}

\begin{figure}[h]
\centering
\includegraphics[width=0.43\textwidth]{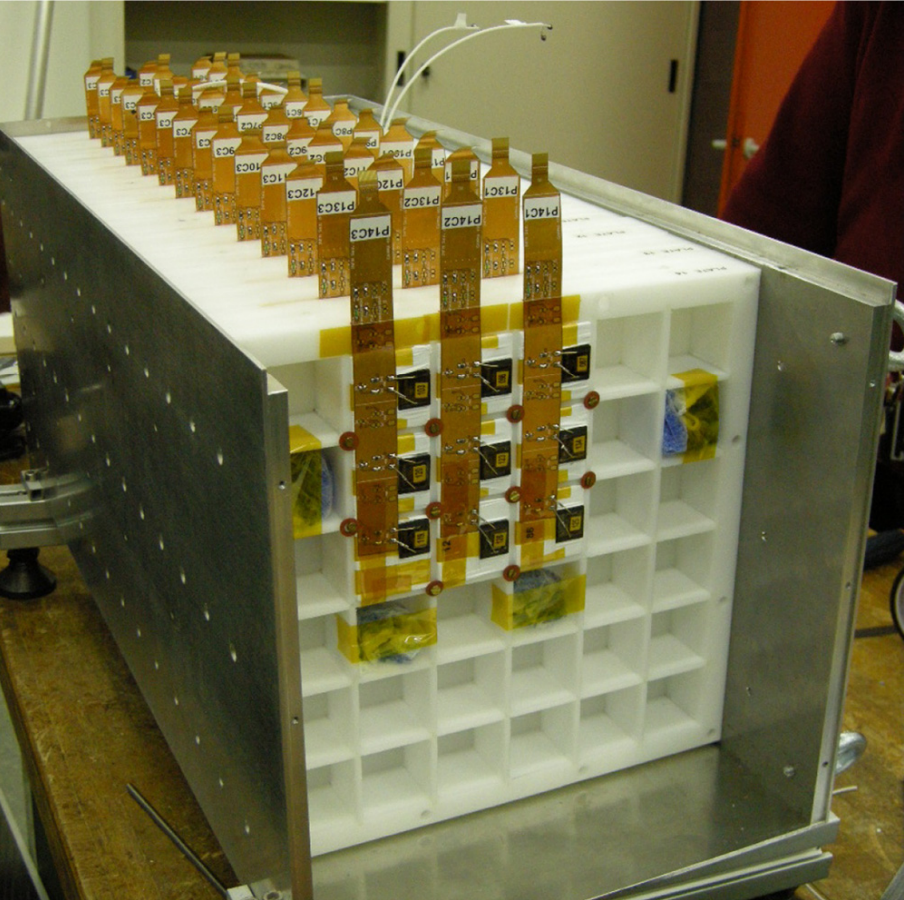}
\caption{The calorimeter prototype inside the Al frame. The photodiodes, as well as the kapton readout cables, are visible on the first layer  (\cite{Mori2013311}).}\label{caloprot}
\end{figure}

The design of the calorimeter is based on a novel configuration. It is composed by 28$\times$28$\times$12 cubes of CsI(Tl), each with a side of 3.6 cm. This particular segmentation permits to reconstruct the shower created by particles coming not only from above but also from the sides of the detector, greatly increasing the geometrical factor (GF) of the instrument. The planar GF is 10.1 m$^2$sr which, taking into consideration the quality cuts necessary to the reconstruction, corresponds to an effective GF of more than 3 m$^2$sr.\\
The expected electron/proton rejection factor is of the order of 10$^5$ with an energy resolution for protons in the 100 GeV-100 TeV energy range, between 30\% and 40\%.\\
The energy resolution for gamma rays, using also the tracker, reaches 1\% at 10 GeV, as shown at the right of fig. \ref{perf}.\\
The possibility of reconstructing the shower of particles coming also from the sides of the detector can be exploited also for gamma-ray observations. Thanks to an angular resolution of the order of some degrees, a more than 2$\pi$ sr field-of-view and a considerable effective area, the GAMMA-400 calorimeter can indeed be used to provide a trigger for observations of transients from the ground.
A prototype of the calorimeter, photograph shown in fig. \ref{caloprot}, has already been tested at the CERN SPS (\cite{Mori2013311}).

\section{CONCLUSIONS}
GAMMA-400 is a space mission dedicated to the study of both gamma rays and cosmic rays, electrons, protons and nuclei. Thanks to the configuration of its detectors it will have an unprecedented energy resolution and an optimal angular resolution. GAMMA-400, considering its performance, will use a multi-messenger approach to search for possible dark matter signal as well as to try solving the remaining issues on the cosmic-rays origin, acceleration and propagation mechanisms. The launch of the satellite is currently scheduled for the beginning of the next decade.

\bigskip 
\bibliography{biblio}

\end{document}